\documentclass[aps,prl,twocolumn,showpacs,superscriptaddress,groupedaddress]{revtex4}  % for review and submission
\usepackage{graphicx}  % needed for figures
\usepackage{dcolumn}   % needed for some tables
\usepackage{bm}        % for math
\usepackage{amssymb}   % for math

\usepackage{amsmath}
\usepackage[utf8]{inputenc}

\newcommand{\dd}{ {\mathrm d} } %% This might require an additional \, too

\newcommand{\exv}[1]{\left\langle{#1}\right\rangle}

\newcommand{\ph}{\varphi}

\newlength{\szovszel}
\newlength{\slashszel} 
\newcommand*{\sls}[1]{\mbox{%
    \settowidth{\szovszel}{\ensuremath{#1}}%
    \settowidth{\slashszel}{\ensuremath{\slash}}%
    \hspace*{0.5\szovszel}%
    \hspace*{-0.5\slashszel}%
    \slash%
    \hspace*{-0.5\szovszel}%
    \hspace*{-0.5\slashszel}%
    \ensuremath{#1}%
  }}

\renewcommand{\d}{\partial}

\begin{document}
%%%%%%%%%%%%%%%%%%%%%%%%%%%%%%%%%%%%%%%%%%%%%%%%%%%%%%%%%%%%%%%%%%%%%%%%%%%%%%%%%%%%%%%%%%%%%%%%%
% The following information is for internal review, please remove them for submission
%\widetext
%\leftline{Version xx as of \today}
%\leftline{Primary authors: Joe E. Physics}
%\leftline{To be submitted to (PRL, PRD-RC, PRD, PLB; choose one.)}
%\leftline{Comment to {\tt d0-run2eb-nnn@fnal.gov} by xxx, yyy}
%\centerline{\em D\O\ INTERNAL DOCUMENT -- NOT FOR PUBLIC DISTRIBUTION}

% the following line is for submission, including submission to the arXiv!!
%\hspace{5.2in} \mbox{Fermilab-Pub-04/xxx-E}

%\title{The FRG Method as a Novel Technique for Calculating \\ Superdense Nuclear Matter Equation of State in Compact Stars}
\title{The Effect of Quantum Fluctuations in the High-Energy Cold Nuclear Equation of State and in Compact Star Observables.}

\author{P\'eter P\'osfay}
\email{posfay.peter@wigner.mta.hu}
\affiliation{Wigner Research Centre for Physics of the H.A.S., P.O. Box, H-1525 Budapest, Hungary}
\affiliation{Institute of Physics, E\"otv\"os Lor\'and University, 1/A P\'azm\'any P. S\'et\'any, H-1117 Budapest, Hungary}
\author{Gergely G\'abor Barnaf\"oldi}
\email{barnafoldi.gergely@wigner.mta.hu}
\affiliation{Wigner Research Centre for Physics of the H.A.S., P.O. Box, H-1525 Budapest, Hungary}
\author{Antal Jakov\'ac}
\email{jakovac@phy.bme.hu}
\affiliation{Institute of Physics, E\"otv\"os Lor\'and University, 1/A P\'azm\'any P. S\'et\'any, H-1117 Budapest, Hungary}

\date{\today}
%%%%%%%%%%%%%%%%%%%%%%%%%%%%%%%%%%%%%%%%%%%%%%%%%%%%%%%%%%%%%%%%%%%%%%%%%%%%%%%%%%%%%%%%%%%%%%%%%%

\begin{abstract}
We present a novel technique to obtain exact equation of state (EoS) by the Functional Renormalization Group (FRG) method, using the expansion of the effective potential in a base of harmonic functions at finite chemical potential. Within this theoretical framework we determined the equation of state and the phase diagram of a simple model of massless fermions coupled to scalars through Yukawa-coupling at the zero-temperature limit.  
We compared our results to the 1-loop and the mean field approximation of the same model and other high-density nuclear matter equation of states. We found a $10-20\%$ difference between these approximations. As an application, we used our exact, FRG-based equation of states to test the effect of the quantum fluctuations in superdense nuclear matter of a compact astrophysical object for the first time. We calculated the mass-radius relation for a compact star using the Tolmann\,--\,Oppenheimer\,--\,Volkov equation and observed a $\sim 5\%$ effect in compact star observables due to quantum fluctuations. 
\end{abstract}

\maketitle
%%%%%%%%%%%%%%%%%%%%%%%%%%%%%%%%%%%%%%%%%%%%%%%%%%%%%%%%%%%%%%%%%%%%%%%%%%%%%%%%%%%%%%%%%%%%%%%%%%%%%%%%%
%\section{Introduction}

Compact astrophysical objects, such as neutron-, quark-, or hybrid stars are the most extreme, natural laboratories for superdense matter. Recently, they become more important by the discovery of gravitational waves~\cite{gw:1,gw:2} since they predicted possible sources of the gravitational radiation.

To explore the inner structure of these {\sl compact stars} is a challenging task due to the lack of direct probes or measurements of their interior. However, spectroscopic radius measurements using X-ray data analysis~\cite{Ozel:2016} and even the gravitational-wave discoveries may provide such strict constraints, which led us to develop more reliable equation of state (EoS) of the superdense matter~\cite{Rezzolla:2016,Walsh:2016}. Thus, modeling the high-density nuclear matter and providing its equation of state is still an open question. 

The success of the above task is shaded by the {\sl masquerade problem}, since different complex and sophisticated EoS result similar behavior and observables of the compact celestial bodies~\cite{Alford:2004}. This motivates us not only to provide perfect EoS, but describe the phase structure of the cold and high-density nuclear matter~\cite{Posfay:2015}. Here we present a new milestone to this study, a nuclear matter EoS, which is consistent with quantum field theory and includes quantum fluctuations as well. 

The calculation of the equation of state in the high-density and zero-temperature limit is usually considered in the mean-field or one-loop approximation. Functional Renormalization Group (FRG) method can extend this description in an exact way, taking into account the effect of quantum fluctuations in the effective action of the system.

In this letter we use the Wetterich-equation to compute the EoS and Litim's regulator is applied regulating the scale dependence~\cite{Litim:2001}. The Local Potential Approximation (LPA) is used to obtain the EoS for the {\sl ansatz} contains a Yukawa-type interaction as described in Ref.~\cite{Barnafoldi:2016}. We present there the calculated EoS which has a Maxwell-construction as an inner nature. The calculated EoS is tested by solving the corresponding Tolman\,--\,Oppenheimer\,--\,Volkov (TOV) equations and investigating the effect of quantum fluctuations via the mass-radius relation, $M(R)$ of compact stars. Comparison of the FRG-based equation of state to other high-density zero-temperature nuclear matter EoS and to the calculated $M(R)$ by various models are given.

%%%%%%%%%%%%%%%%%%%%%%%%%%%%%%%%%%%%%%%%%%%%%%%%%%%%%%%%%%%%%%%%%%%%%%%%%%%%%%%%%%%%%%%%%%%%%%%%%%%%%%%%%%%%%%%%%%

%\section{The FRG Method for a Yukawa-type Model}

The functional renormalization group method is a general way to find the effective action of a system. This formalism led us to calculate low-energy effective (observable) quantities by gradual momentum integration of a theory defined at some high-energy scale, $k$. Since low-scale effective quantities incorporate quantum fluctuations using FRG at finite temperature, one may calculate the equation of state of the system including the quantum fluctuations as well. 

Within the FRG framework the quantum $n$-point correlation function is calculated by the gradual path integration. This can be achieved by introducing a regulator term, $R_{k}$ in the generator functional, $Z_{k}[J]$, which acts as a mass term and suppress modes below scale, $k$ as explained in Refs.~\cite{Wetterich:1989xg,Gies:2006wv}. Thanks to this regulator term, the effective action becomes scale-dependent, which scale dependence is given by the Wetterich-equation~\cite{Wetterich:1992yh} 
\begin{equation} 
\partial_{k} \Gamma_{k}=\frac{1}{2} \, \int \dd p^D \,  {\rm STr} \, \left [ \left( \partial_{k}R_{k} \right) \left( \Gamma_{k}^{(2)} + R_{k} \right)^{-1} \right],
\label{wetterich}
\end{equation}
where $\Gamma_{k}^{(2)}$ is the second derivative matrix of the effective action. The term 'STr' is stand for the normal \textit{trace} operation but includes a negative sign for fermionic fields and sums over all indices. The low-scale (observable) effective action is computed by integrating the Wetterich-equation~\eqref{wetterich}, from the classical limit at some UV-scale $k=\Lambda$ to the IR-scale $k=0$, where quantum effects are taken into account. The initial condition in this integration is the UV-scale (classical) action $\Gamma_{k=\Lambda}$, which has to be chosen in a way, that the low-scale effective action reproduces physical quantities, correctly. 
%
%\begin{figure}[!h]
%\begin{center}
%\includegraphics[width=0.20\textwidth]{./figs/fermionloop.png}
%\end{center}
%\vspace*{-0.5truecm}
%\caption{The Wetterich-equation including the exact propagator and the regulator.}
%\label{wetter-f}
%\end{figure}

Here, we use a simple Yukawa-type model with one bosonic and one fermionic degree of freedom described by the bare action. This is defined at scale $\Lambda$,
\begin{eqnarray}
  && \Gamma_\Lambda[\ph,\psi] =  \nonumber \\
  && = \! \int \dd^4x
    \left[\bar \psi(i\sls \d - g_0\ph)\psi + \frac12 (\d_\mu\ph)^2-\frac{m_0^2}2\ph^2-\frac{\lambda_0}{24}\ph^4
    \right]. \nonumber \\
\end{eqnarray}
As we described in Ref.~\cite{Barnafoldi:2016} this model has two phases: (i) in the symmetric phase the fermion is massless, (ii) in the Spontaneous Symmetry Breaking (SSB) phase the fermion mass is $g\exv{\ph}$.

\begin{figure}[!h]
\begin{center}
\includegraphics[width=0.30\textwidth]{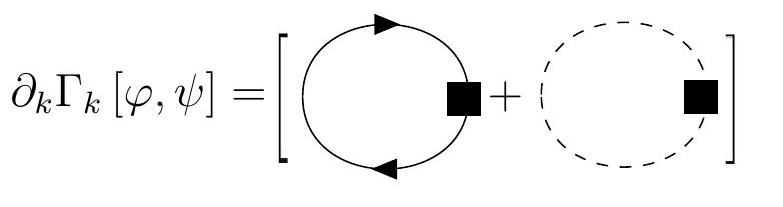}
\end{center}
%\vspace*{-0.5truecm}
\caption{The Wetterich-equation with our {\sl ansatz} including a Yukawa-like coupling with one fermion and one boson field.}
\label{fig:wetter-fb}
\end{figure}

To treat this model with the FRG method we need an {\sl ansatz} for the effective action at scale $k$. We choose the simplest possible one, where only the bosonic effective potential depends on the scale:
\begin{equation}\label{eq:wetter-fb}
  \Gamma_k[\ph,\psi] = \int \! \dd^4x\left[\bar \psi(i\sls \d -g\ph)\psi +
    \frac12 (\d_\mu\ph)^2 - U_k(\ph)\right].
\end{equation}
Note, neither wave function renormalization, nor the running of the Yukawa coupling are taken into account however, both effects can be easily adapted into the present method. This {\sl ansatz} represented by Fig.~\ref{fig:wetter-fb}.

The integrated Wetterich-equation~\eqref{wetterich} for this model can be rewritten after applying the three-dimensional Litim regulator at finite temperature $T$ and at finite chemical potential $\mu$,
\begin{eqnarray}\label{wetterich2}
 && \d_k U_k = \frac12\mathop{\mathrm{STr}} \ln \left[ R_k + \Gamma_k^{(2)} \right] = \frac{k^4}{12\pi^2} \times \nonumber \\
 && \times  \left[
    \frac{1+2n_B(\omega_B)}{\omega_B} -
    4\frac{1-n_F(\omega_F-\mu)-n_F(\omega_F+\mu)}{\omega_F}\right], \nonumber \\
\end{eqnarray}
where $n_B(\omega)$ and $n_F(\omega)$ are the Bose\,--\,Einstein and the Fermi\,--\,Dirac
distributions, respectively
\begin{equation}
  n_{B/F}(\omega)= \frac1{1\mp e^{-\beta\omega_{B/F}}}, 
\end{equation}
with $ \beta=\frac1T$ and fermi and bose states are
\begin{equation}
 \omega_B^2 = k^2 +\d_\ph^2 U \qquad \textrm{and} \qquad \omega_F^2 = k^2 +g^2\ph^2.
\end{equation}
In the case of $T=0$ and $\mu>0$ the Bose\,--\,Einstein distribution does not give contribution, but the Fermi\,--\,Dirac distribution reduces $n_F(\omega)$ to $\Theta(-\omega)$ and this simplifies equation~\eqref{wetterich2} to:
\begin{equation}
  \d_k U_k = \frac{k^4}{12\pi^2}\left[\frac1{\omega_B} -
    4\frac{\Theta(\omega_F-\mu)}{\omega_F}\right] .
\end{equation}
with the initial condition  
\begin{equation}
    U_{\Lambda} (\ph) = \frac{m_0^2}2\ph^2+\frac{\lambda_0}{24}\ph^4 \ .
\end{equation}
The presence of the step function, $\Theta(\omega)$ generates two different domains, where two different differential equations evolve the potential in $k$. The boundary of these domains is called Fermi-surface $S_F$, which can be determined from requiring $ \omega_F(k,\ph)|_{S_F} = \mu $.
The surface can be characterized either by $k=k_F(\ph)$ or by $\ph=\ph_F(k)$. In our case these read as
\begin{equation}
  k_F=\sqrt{\mu^2 - g^2\ph^2} \qquad  \textrm{and} \qquad \ph_F=g^{-1} \sqrt{\mu^2-k^2}.
\end{equation}
The surface $S_F$, in terms of $k$ and $g\ph$, is a circle with radius $\mu$, and for $\mu=0$ it disappears. The Fermi-surface divides the coordinate space into two parts; we will denote the high energy regime by ${\cal D}_>$, the low energy regime by ${\cal D}_<$, where the following differential equations hold:
\begin{description}
\item if $ (k,\ph)\in {\cal D}_>:= \left\{ (k,\ph) \,|\, k^2 +g^2\ph^2 > \mu^2\right\}$, then
\begin{equation}
    \label{eq:Ua}
  \d_k U_k = \displaystyle\frac{k^4}{12\pi^2}\left[\frac1{\omega_B} - \frac{4}{\omega_F}\right] ;
\end{equation}
\item if $ (k,\ph)\in {\cal D}_<: = \left\{ (k,\ph) \,|\, k^2 +g^2\ph^2 < \mu^2\right\}$, then
\begin{equation}
    \label{eq:Ub}
  \d_k U_k = \displaystyle\frac{k^4}{12\pi^2}\frac1{\omega_B} .
\end{equation}
\end{description}
and the solution must be continuous at $k=k_F$.

For the solution we need to use standard FRG techniques, e.g. with discretization or with polynom expansion. 
The value of the potential at the boundary can be determined by cutting out the Fermi-surface from the zero chemical potential solution. In Ref.~\cite{Barnafoldi:2016} we introduced a coordinate transformation which maps the circle-like Fermi-surface to a rectangular one, while keeps the symmetries of the differential equations. Applying this circle-to-rectangle transformation and a harmonic expansion the Wetterich-equation can be solved numerically. 

Since exact solution can be given in the mean field approximation, we took $v=f_{\pi}$, $g=m_N/v$, 
and $\lambda =3 m_{\sigma}^2/v^2$ with the values $m_N=m_{\sigma}=0.938$ GeV and $f_{\pi}=0.093$ GeV. We choose the chemical potential %,$\mu_{MF}$, 
close to the value of the first-order phase transition $\mu_{MF} \approx 0.6177 m_N$.

\newpage
%%%%%%%%%%%%%%%%%%%%%%%%%%%%%%%%%%%%%%%%%%%%%%%%%%%%%%%%%%%%%%%%%%%%%%%%%%%%%%%%%%%%%%%%%%%%%%%%%%%%%%%%%%%%%%%%
%\section{Results: Equation of State and Mass-radius Relation of Compact Stars}
\begin{figure}[!h]
\begin{center}
\includegraphics[width=0.48\textwidth]{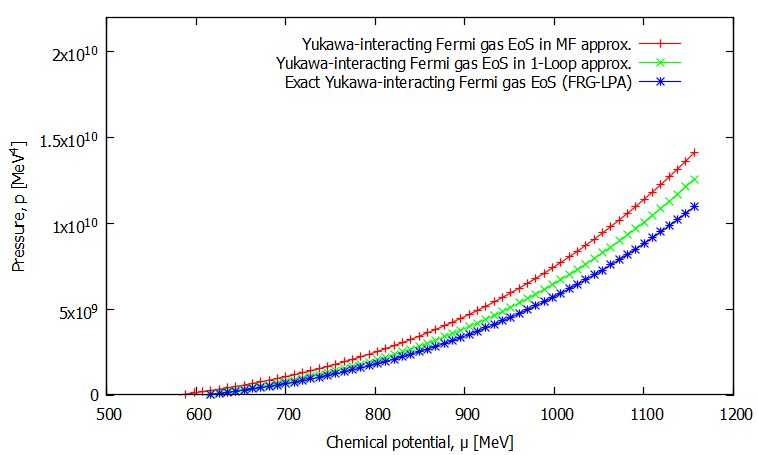}
\end{center}
\caption{ Equation of state, $p(\mu)$ calculated from the functional renormalization group method (FRG-LPA, {\sl blue '$\ast$'}), in the 1-loop (1-Loop, {\sl green '$\times$'}), and in the mean field approximation (MF, {red '$+$'}) from the bottom to the top respectively.}
\label{frgeos}
\end{figure}

As a result, we provided a novel potential solution by the FRG method. This converge fast where the potential is convex, in contrary, where it is concave converges slowly, but this can be treated easily, because it generates a native Maxwell-construction. Comparison of the FRG-based EoS results to the mean-field and one-loop approximation of the same Yukawa-like case is plotted on Fig.~\ref{frgeos} as $p(\mu)$. Note, here we are in the zero temperature limit with finite chemical potential. One can see, that at fixed $\mu$ values, the FRG-based EoS is the softest (FRG-LPA, {\sl blue '$\ast$'}), followed by the stiffer one-loop (1-Loop, {\sl green '$\times$'}) and finally the stiffest mean field (MF, {red '$+$'}) approximation result.   

As Fig.~\ref{frgeos} presents, at fix $\mu$ values, the 1-loop approximation and FRG-based calculations are closer to each other than the mean field one. The calculated pressures $p_i(\varepsilon)$ in the 1-loop and mean field approximations normalized by the FRG-based $p_{FRG}(\varepsilon)$ for the same energy density values: $(p_{1L}/p_{FRG})(\varepsilon) $ shows $10\%$ extra stiffness, while $ (p_{MF}/p_{FRG})(\varepsilon) $ is $20\%$ stiffer. Deviation have a slight increase as higher the energy density, which is due to the weak ($\sim \log E$) energy dependence of the coupling, which certainly depends on the order of the approximation.

\begin{figure}[!h]
\begin{center}
\includegraphics[width=0.48\textwidth]{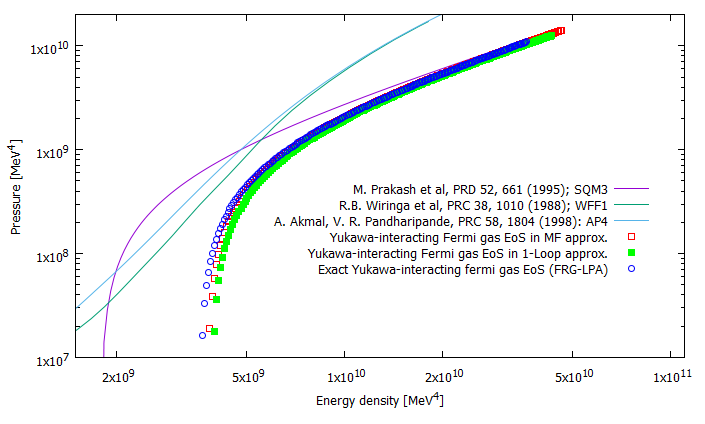}
\end{center}
\caption{Comparison of the calculated EoS on logarithmic scales in the mean field ({\sl red open squares}) and 1-loop ({\sl green full squares}) approximation and via the FRG ({\sl blue open circles}). Results are compared to SQM3~\cite{SQM3}, WFF1~\cite{WFF1} and AP4~\cite{AP4} EoS from Ref.~\cite{Ozel:2016}.}
\label{eos-mr}
\end{figure}

Apart from the plainnes of Yukawa-like model, we compared our FRG-based EoS results to some other EoS~\cite{SQM3,WFF1,AP4} taken from Ref.~\cite{Ozel:2016}, which are typically used in compact star models. On Fig.~\ref{eos-mr}, we found, our EoS calculations are the closest to the model prediction, denoted by 'SQM3', especially at the highest energy density values. On the other hand, our FRG-based EoS turn down at a twice higher, $\varepsilon \approx 4 \cdot 10^{9}$ MeV\textsuperscript{4} than the SQM3~\cite{SQM3}. Calculating $p(\varepsilon)$, evolution of the different approximations are slightly varies due to the considered orders of the couplings and terms.

\begin{figure}[!h]
\begin{center}
\includegraphics[width=0.49\textwidth]{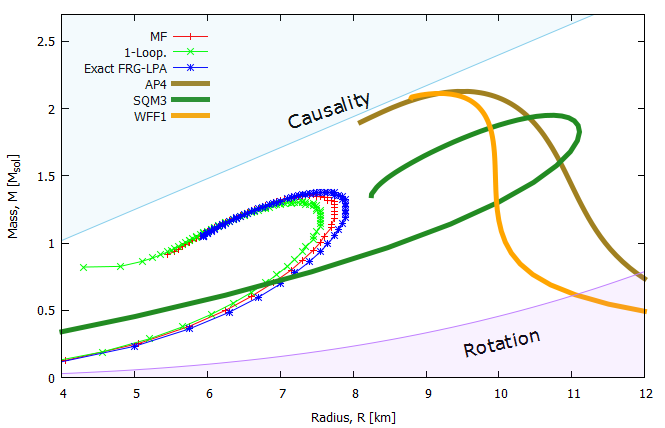}
\end{center}
\caption{\label{eos-mr-ozel} Comparison of the $M(R)$ diagrams for compact stars calculated by the equation of states with our Yukawa-like coupling in the mean field and 1-loop approximation to the FRG results. The SQM3~\cite{SQM3}, WFF1~\cite{WFF1} and AP4~\cite{AP4} EoS from Ref.~\cite{Ozel:2016} are drawn jointly.}
\end{figure}

Fig.~\ref{eos-mr-ozel} presents the calculated mass-radius diagram $M(R)$, calculated by the Tolman\,--\,Oppenheimer\,--\,Volkov equation for a static, spherical symmetric, one fermion component compact star with our Yukawa-like interaction. We compared the $M(R)$ results of the exact FRG EoS, the 1-loop-, and the mean-field approximations. In general one can observe this simple, Yukawa-like model provide surprisingly realistic $M(R)$ curves within the $M \lesssim 1.4 M_{\odot}$ and $R\lesssim 8$ km region although, never reach the observed compact star mass $\sim 2M_{\odot}$.  %%% Maybe ref need to added 

According to the EoS results presented on Fig.~\ref{frgeos}, the mass-radius relations are converges  both at high and low energy-density limit for all EoS cases. The stiffness difference is reflected in the results: the softest, exact FRG forms massive, $M_{max} \approx 1.38 M_{\odot}$ and larger, $R_{max} \approx 7.7$ km compact star, while the stiffest mean field and 1-loop approximated curves are $\sim 5\%$ less smaller: less massive with about $0.1 M_{\odot}$ and smaller with {\sl cc}. half km. This result is quite remarkable, since the exact solution provided by the FRG-method accounted the quantum fluctuations, which matters both in sense of the EoS and the final observables of the compact star. 

Calculated EoS cases were drawn together with other model predictions from Ref.~\cite{Ozel:2016} for comparison on Fig.~\ref{eos-mr-ozel}. Mass-radius relation calculated by SQM3 and our exact FRG-method overlaps nicely at the high-$p$ and high-$\varepsilon$ regime, where the most compact (smaller) stars are. Similarly, maximum masses and radii of our models with Yukawa-like coupling are well below the SQM3 results due to the higher cutoff in the $p(\varepsilon)$.  

%%%%%%%%%%%%%%%%%%%%%%%%%%%%%%%%%%%%%%%%%%%%%%%%%%%%%%%%%%%%%%%%%%%%%%%%%%%%%%%%%%%%%%%%%%%%%%%%%%%%%%%%%%%%%%%%%%
%\section{Summary}

Summary: using a novel technique presented in Ref.~\cite{Barnafoldi:2016} we calculated the equation of state at the zero-temperature limit and at finite chemical potential for a simple model with massless fermions which are coupled to scalars through Yukawa-coupling. Calculations were made in the framework of the functional renormalization group method, using the Wetterich-equation. We compared our {\sl exact} results to the same Yukawa-like model derived in the 1-loop and mean filed approximations. We found, quantum fluctuations matter, and accounting them provides the softest $p(\mu)$ equation of state in the exact FRG-based solution. Matching our model to the earlier equation of states, we have found consistency, although the plainness of the Yukawa-like model. 

In this Letter we used our exact FRG-based equation of state to test the effect of quantum fluctuations in superdense nuclear matter inside compact astrophysical objects. We could successfully validate our concept by calculating the mass-radius relation for a compact star and compare to other results from Ref.~\cite{Ozel:2016}. Although our simple Yukawa-like model predicted smaller but consistent compact stars with $M \lesssim 1.4M_{\odot}$ and $R \lesssim 8$ km, a $5\%$ difference were observed between the exact FRG-based equation of state and its 1-loop or mean field approximations. 

The obtained result is quite remarkable, since the exact solution provided by the FRG-method took into account the quantum fluctuations, which matters both in sense of the EoS ($10-20\%$) and finally in the observables of the compact star ($5\%$). We suppose, taking into account the interaction in a realistic e.g. Walecka-like model this effect is expected to be more stronger.

%%%%%%%%%%%%%%%%%%%%%%%%%%%%%%%%%%%%%%%%%%%%%%%%%%%%%%%%%%%%%%%%%%%%%%%%%%%%%%%%%%%%%%%%%%%%%%%%

\section*{Acknowledgments}

This work is supported by the Hungarian Research Fund (OTKA) under contracts No. K104292, K104260, NK106119, K120660, NKM-81/2016 MTA-UA bilateral mobility program, NIH TET\_12\_CN-1-2012-0016, and NewCompStar COST Action MP1304. Author GGB also thanks the J\'anos Bolyai Research Scholarship of the Hungarian Academy of Sciences. 

%%%%%%%%%%%%%%%%%%%%%%%%%%%%%%%%%%%%%%%%%%%%%%%%%%%%%%%%%%%%%%%%%%%%%%%%%%%%%%%%%%%%%%%%%%%%%%%%

\end{document}